# Single-Crystal Growth and Characterization of Cuprate Superconductor (Hg,Re)Ba$_2$Ca$_2$Cu$_3$O$_{8+\delta}$

Yutaro Mino[1,2], Shigeyuki Ishida[2*], Junichiro Kato[1,2], Shungo Nakagawa[2,3], Takanari Kashiwagi[3], Takahiro Nozue[1,2], Nao Takeshita[2], Kunihiro Kihou[2], Chul-Ho Lee[2], Taichiro Nishio[1†], Hiroshi. Eisaki[2]

[1]*Tokyo University of Science, Tokyo 162-8601 Japan*
[2]*National Institute of Advanced Industrial and Science Technology (AIST), Ibaraki 305-8568 Japan*
[3]*University of Tsukuba, Ibaraki 305-8577 Japan*

We grew (Hg,Re)Ba$_2$Ca$_2$Cu$_3$O$_{8+\delta}$ ((Hg,Re)1223) single crystals with good reproducibility via the single-step flux method using monoxides as raw materials. A double-sealing method using a thick-walled quartz tube and a stainless-steel container was adopted for explosion protection. The maximum crystal size was approximately 1 mm × 1 mm in the *ab* plane and 0.04 mm in thickness. The crystal was square-shaped, reflecting the tetragonal crystal structure of (Hg,Re)1223. Magnetic susceptibility measurements indicated a critical temperature of 130 K. The in-plane resistivity exhibited a linear temperature dependence, indicating that the sample was close to optimal doping level. The out-of-plane resistivity was also measured, and the anisotropy parameter was 250–650 at 300 K.

## 1. Introduction

More than 30 years have passed since the discovery of high-critical-temperature (high-$T_c$) cuprates.[1] They continue to exhibit the highest $T_c$ values at the ambient pressure among existing superconductors, making them the most promising materials for superconductivity applications. Although numerous studies have been conducted, the mechanism of high-$T_c$ superconductivity remains to be elucidated, and various theoretical and experimental studies are in progress. One approach for clarifying the high-$T_c$ mechanism is to investigate the physical properties of the material with the highest $T_c$ and determine the essential parameters for realizing a high $T_c$. Among cuprates, the Hg-based HgBa$_2$Ca$_{n-1}$Cu$_n$O$_{2n+2+\delta}$ (Hg12($n$-1)$n$) with $n$ = 3, i.e., Hg1223,[2] has the highest $T_c$, which exceeds 130 K at the ambient pressure and reaches 150 K under a pressure of 30 GPa.[3,4] The reasons for Hg1223 having the highest $T_c$ are thought to be as follows.

(1) The number of CuO$_2$ planes in a unit cell is 3, which is empirically the optimal number.[5,6]

(2) The crystal structure is tetragonal with high symmetry.[7]

(3) The CuO$_2$ planes are flatter than those of the other cuprates.[8]

(4) The dopant impurities (disorders) are located away from the CuO$_2$ planes.[9]

However, owing to the lack of high-quality large single crystals of Hg1223, information is lacking on the physical properties of Hg1223, such as the anisotropy parameter, superconducting gap structure, Fermi surface shape, band dispersions, charge dynamics, spin excitation spectra, phonon spectra, and their carrier concentration dependences. The comparison of these properties with those of other representative cuprates with lower $T_c$, such as La$_{2-x}$Sr$_x$CuO$_4$, YBa$_2$Cu$_3$O$_{7-\delta}$, and Bi$_2$Sr$_2$CaCu$_2$O$_{8+\delta}$, will help elucidate the key ingredients for high $T_c$ in Hg1223.

Table 1 summarizes previous reports on the growth of Hg1223 single crystals.[10-16] The key points of single-crystal growth in previous studies are listed below.

(1) *Vaper pressure control*: The sample should be encapsulated in a container to control the high vapor pressure of Hg. Typically, the sample is sealed in an evacuated quartz tube with an inner diameter of 10–13 mm, thickness of 2–3 mm, and length of 70–90 mm. For explosion protection, the quartz tube may be double-sealed with a larger quartz tube or a metal container. In the latter case, the inside of the metal container is filled with N$_2$ gas, which provides a counterpressure to prevent the quartz tube from exploding.

(2) *Starting materials*: Various combinations of starting materials have been reported. In several studies,[10,12,15] oxide materials (HgO, BaO, CaO, and CuO) were used as raw materials to directly grow Hg1223. In other studies,[11,13,14,16] carbonates or nitrates (BaCO$_3$, Ba(NO$_3$)$_2$, and CaCO$_3$, which are stable in air) were first reacted with CuO to form a precursor, which was then mixed with HgO to grow Hg1223.

(3) *Chemical substitution to Hg*: In several studies,[11-14] high-valence metals such as Pb, Bi, and Re are doped into the Hg site because the partial replacement of Hg with high-valance metals makes the Hg1223 phase chemically stable.

(4) *Material handling*: To ensure the safe handling of toxic HgO and to prevent the degradation of water-absorbing BaO, CaO, and Ba-Ca-Cu-O precursors, they should be handled in a glovebox filled with dry N$_2$ or Ar gas.

(5) *Crucible material*: Alumina is often used as a crucible material. However, Ueda et al.[14] and Wang et al.[16] reported that Al and Y can replace Ca and Cu sites; therefore, BaZrO$_3$ and ZrO$_2$ are preferred as crucible materials.

(6) *Heat-treatment conditions*: The optimal temperature profiles depend on various

conditions, such as the starting materials, composition ratio, and flux agent. For example, the maximum temperature varies widely, from 800 to 1150 °C; the Au amalgam method allows growth at a lower temperature (800 °C),[10] while the high-pressure method requires a higher temperature of 1040–1150 °C.[12,16] In the method of Ueda et al.,[14] $BaF_2$ is added to lower the melting point and promote crystal growth.

In this study, we mainly focus on three key factors to grow large Hg1223 single crystals with high reproducibility. The first is the choice of starting materials. It has been reported that the precursors have uncertainties in their chemical compositions.[17,18] Such uncertainties make it difficult to control the starting composition ratio, likely reducing the reproducibility.[16,17] Therefore, with regard to reproducibility, single-step synthesis using oxide materials with less uncertainties in their chemical compositions is preferable.

The second factor is the starting composition ratio. Ueda et al.[14] reported that Hg12(n-1)n with different $CuO_2$ plane numbers ($n$ = 2–4) can be formed by changing the starting composition ratio. In addition, Lin et al.[13] reported that $Ba_2Cu_3O_5$ becomes the main phase when CuO is enriched. Furthermore, Meng et al.[17] pointed out that the Hg12(n-1)n phases can be efficiently synthesized by adjusting the Hg vapor pressure. Then, Re substitution for Hg, which increases the chemical stability of the Hg1223 phase and reduce the Hg vapor pressure, is expected to promote the crystal growth of Hg1223. Thus, it is important to investigate the optimal BaO-CaO-CuO and Hg-Re ratios.

The third factor is the heat-treatment conditions. As shown in Table 1, the optimal temperature profiles (maximum temperature, holding time, cooling rate, and final temperature) differ significantly among the reports. Therefore, to obtain large Hg1223 single crystals, it is essential to optimize the temperature profile depending on the starting composition.

Accordingly, we attempted crystal growth more than 100 times and established a method for growing (Hg,Re)1223 single crystals using the self-flux method by optimizing the growth conditions. Under the optimal conditions, large single crystals with sizes up to 1 × 1 × 0.04 $mm^3$ were obtained, and crystals with sizes of approximately 0.8 × 0.8 $mm^2$ in *ab*-plane area were obtained with good reproducibility. The physical properties of several samples were investigated, confirming that the sample dependence was small. Details regarding the experimental method are presented in Section 2. In Section 3, the crystal growth conditions are discussed. In Section 4, the crystallinity and chemical compositions of the obtained crystals are described. Lastly, the superconducting and normal-state properties of the crystals are presented in Section 5.

## 2. Experimental Section

*2.1 Synthesis method*

To synthesize (Hg,Re)1223 in a single step, monoxides HgO (Alfa Aesar, 99.99%), $ReO_3$ (Thermo Scientific, 99.9%), BaO, CaO, and CuO (Furuuchi Chemical, 99.9%) were used as raw materials. Additionally, $BaF_2$ (RARE METALLIC, 99.99%) was added to lower the melting point. BaO was obtained by calcining $BaO_2$ (Furuuchi Chemical, 99.9%) under vacuum. Here, $BaCO_3$ was not used, because of the possibility that carbonate remains in BaO after calcining, which disturbs the formation of the Hg1223 phase.[17,19] CaO was obtained by calcining $CaCO_3$ (Furuuchi Chemical Corporation, 99.99%) at 1200 °C in air for 1 d. No reports suggest that carbonate remains in CaO, even when it is made from $CaCO_3$. To improve reproducibility, the raw materials were handled in a high-performance glovebox with gas purification ($O_2$, $H_2O$ < 1 ppm). BaO was handled with special care to avoid the formation of impurities because it easily reacts with $CO_2$ and $H_2O$ in air to form $BaCO_3$ and $Ba(OH)_2$. The raw materials with a total weight of 4.5–6.0 g were mixed in a mortar and pelletized into 8-mm-diameter pellets.

A $ZrO_2$ crucible (9.5 mm inner diameter (ID) × 12 mm outer diameter (OD) × 50 mm height) was used to avoid contamination with Al or Y. Quartz wool was placed at top of the crucible to prevent HgO from evaporating because of radiant heat during the quartz-tube sealing process. The crucible was then placed in a thick-walled quartz tube (14 mm ID × 19 mm OD). Quartz wool was also placed at the bottom of the quartz tube as a cushion to avoid damage to the bottom due to the impact caused by the insertion of the crucible into the quartz tube. The quartz tube was capped to prevent the sample from being exposed to air, removed from the glovebox, and sealed. The length of the quartz tube was kept in the range of 75 ± 3 mm to control the effect of the internal pressure.

For the explosion protection, the encapsulated quartz tube was placed in a stainless-steel container and sealed in the glovebox. By sealing the quartz tube in a stainless-steel container filled with $N_2$, a counterpressure of approximately 4 atm was provided to the quartz tube around the growth temperature (~1000 °C). This prevented the quartz tube from breaking. Moreover, the stainless-steel container prevented leakage of Hg, even when the quartz tube exploded. The stainless-steel container was placed in an alumina crucible and the surrounding space was filled with alumina powder to avoid oxidation of the stainless steel. A schematic of the setup and photographs are shown in Figs. 1(a) and 1(b), respectively.

The growth set was placed in a box-type electric furnace, heated to 800 °C in 6 h, and held at 800 °C for 5 h. During this process, (Hg,Re)1223 polycrystals were expected to form, resulting in a reduction in the Hg pressure in the quartz tube at higher temperatures.[16] The temperature was then increased to 960–980 °C in 3 h, held for 0–10 h, and then slowly cooled to 900–940 °C at a rate of 0.33–2 °C/h. The effects of the starting temperature, holding time, cooling rate, and final temperature on crystal growth are discussed in detail in Section 3.5. After growth, the stainless-steel container was cut and the quartz tube was broken to remove the crucible. The crucible was crushed using a press machine to separate the single crystals.

*2.2 Sample characterization*

The surfaces of the obtained crystals were examined using scanning electron microscopy (SEM, Hitachi TM3000), and the chemical compositions were investigated using energy-dispersive X-ray spectroscopy (EDX, Oxford Instruments SwiftED3000). To evaluate the crystallinity, the 00$l$ peaks were examined using X-ray diffraction (XRD, Rigaku Ultima VI) and back-reflection Laue images were obtained using an XRD instrument (Rigaku Photonic Science). Magnetization measurements were performed using a SQUID magnetometer equipped with a superconducting magnet (Quantum Design MPMS-XL and MPMS-3). The *ab*-plane/*c*-axis resistivity was measured via the 4-probe or 6-probe method using a physical property measurement system (Quantum Design PPMS). For electrical resistivity measurements, the electrodes were drawn by hand using Au paste (Tanaka Kikinzoku Kogyo TR-1404) and sintered at 600 °C for 12 h to achieve a contact resistance of <3 Ω.

## 3. Results of Crystal Growth

As mentioned in Section 1 various parameters affect the single-crystal growth of (Hg,Re)1223. The optimal conditions in this study were determined through >100 trials by repeating the growth and characterization of the crystals and adjusting the starting composition and heat-treatment profile. After introducing the optimal conditions in Section 3.1, we describe how each factor (BaO-CaO-CuO ratio, Hg/Re ratio, charge amount, and heat-treatment conditions) affected the crystal growth.

*3.1 Summary of optimal conditions*

The optimal molar ratio of starting composition obtained in this study was $HgO:ReO_3:BaO:CuO:CaO:BaF_2$ = 0.8:0.2:2.4:1.2:3.1:0.4. Figure 2 shows the optimal temperature profile. The temperature was increased to 800 °C in 6 h, held for 5 h, increased to 980 °C in 3 h, held for 5 h, and reduced to 920 °C at 0.5 °C/h, followed by furnace cooling. By employing the above conditions, single crystals with sizes of 0.8 × 0.8 × 0.03 mm$^3$ were reproducibly obtained. The maximum dimensions were 1 × 1 × 0.04 mm$^3$.

*3.2 Dependence on Ba-Ca-Cu ratio*

Figs. 3(a) and 3(b) show the triangular phase diagram of BaO-CaO-CuO and a magnified view of the area surrounded by the red triangle in Fig. 3(a), respectively. Here, the stoichiometric compositions are marked by a square (■) for Hg1212, a diamond (◆) for Hg1223, and a triangle (▲) for Hg1234. The circles (●) indicate compositions for which (Hg,Re)1223 single crystals with sizes of 0.5 × 0.5 mm$^2$ or larger were obtained, and crosses (×) indicate compositions for which crystals of such sizes were not obtained. The blue circle corresponds to the optimal composition in this study (BaO:CaO:CuO = 2.4:1.2:3.1).

Starting from the composition closest to the Hg1223 stoichiometry (yellow cross, BaO:CaO:CuO = 1.8:1.5:2.5), (Hg,Re)1223 crystals smaller than 0.1 × 0.1 mm$^2$ were obtained. When the ratio of Ca relative to Ba and Cu was reduced (gray cross, Ba:Ca:Cu = 2.4:1.5:3.1), slightly larger crystals with sizes of approximately 0.1 × 0.1 mm$^2$ were obtained. The ratio of Ca was further reduced along the dashed line, which corresponds to Ba:Cu = 2.4:3.1. At the orange circle (Ba:Ca:Cu = 2.4:1.0:3.1), where Ca was nearly half of the stoichiometry, relatively large crystals with sizes of approximately 0.5 × 0.5 mm$^2$ were obtained. With a further reduction in the Ca ratio in this direction, as indicated by the purple cross (Ba:Ca:Cu = 2.4:0.7:3.1), no single crystals were obtained, and Ba$_2$Cu$_3$O$_5$ was the major phase. Thus, the composition range in which large crystals can be obtained is limited, and the formation of (Hg,Re)1223 depended on the ratio of Ca relative to Ba and Cu. The optimal Ca ratio was approximately half of the (Hg,Re)1223 stoichiometry, suggesting that Ba-Cu-O acts as self-flux and promotes crystal growth, which is consistent with a previous report.[13]

The composition indicated by the green circle (Ba:Ca:Cu = 2.3:1.4:3.3) yielded the (Hg,Re)1234 phase in addition to (Hg,Re)1223, as revealed by SEM/EDX observations and magnetization measurements. The results suggest that (Hg,Re)1234 single crystals can be grown by slightly increasing the Ca and Cu ratios relative to Ba in the starting composition.

*3.3 Effect of changing Hg:Re ratio*

The Hg:Re ratio also affects the crystal growth of (Hg,Re)1223. We attempted to grow Re-free Hg1223 using the same setup, which resulted in explosion of the quartz tube. This supports the notion that the addition of Re suppresses the vapor pressure of Hg. When Hg:Re = 0.9:0.1 was employed, the largest crystal was smaller than 0.5 × 0.5 mm$^2$. Compositional analysis via SEM/EDX revealed that the Hg:Re ratio of the grown crystals was 0.78:0.22, indicating that the amount of Re in the crystal exceeded that in the starting composition, which possibly stabilized the (Hg,Re)1223 phase. In contrast, the starting composition of Hg:Re = 0.7:0.3 resulted in the crystals with Hg:Re = 0.77:0.23, suggesting that the stable Re ratio in (Hg,Re)1223 is fixed at a value close to 0.2. Thus, the optimal Hg:Re ratio in this study was determined to be 0.8:0.2.

*3.4 Effect of charging amount*

In general, increasing the amount of starting materials should increase the number and volume of grown crystals. However, in the crystal growth of Hg1223, as the total amount of materials increases, the Hg vapor pressure increases, increasing the risk of explosion. The optimal amount used in this study was 6.0 g. When the total amount was reduced to 4.5 g, the crystal size did not change significantly, while the number of crystals decreased. When the amount was increased to 9 g, the molten samples came out of the crucible and reacted with the quartz wool and tube, which significantly disturbed the crystal growth, although the quartz tube did not explode.

*3.5 Effect of heat-treatment profile*

The optimal heat-treatment profile is shown in Fig. 2. Here, we describe how the growth is affected by the crystal-growth parameters: (a) maximum temperature, (b) holding time, (c) cooling rate, and (d) final temperature.

The molten flux reacts with $ZrO_2$ crucibles at high temperatures of approximately 1000 °C. Accordingly, the maximum temperature should be as low as possible but higher than the melting temperatures of the components. When the maximum temperature was 960 °C, the sample did not melt, and no crystal was obtained. When the maximum temperature was increased to 970 °C, the sample pellet melted, and crystals with a size comparable to those for 980 °C were obtained. However, the pellet sometimes did not melt completely even when the same temperature was employed, suggesting that the melting point of the sample was close to 970 °C. We found that 980 °C is the optimal temperature for the crystal growth of (Hg,Re)1223, because the sample melted reproducibly and did not react significantly with the crucible.

The typical size of crystals grown without holding time was approximately $0.5 \times 0.5$ mm$^2$. When the holding time was 3 h, crystals with a maximum size of $0.9 \times 0.9$ mm$^2$ were obtained, which was nearly the largest size in this study. When the holding time was extended to 10 h, the quartz tube exploded, possibly because it could not withstand the high pressure of Hg vapor for a long time. According to these results, the optimal holding time was determined to be 5 h.

We found that the optimal cooling rate was 0.5 °C/h. When the cooling rate was increased to

1 °C/h, the crystal size decreased to less than 0.5 × 0.5 mm². Additionally, when the cooling rate was decreased to 0.33 °C/h, the crystal size decreased to less than 0.3 × 0.3 mm². In this case, the discoloration of the crucible surface was more noticeable. This indicates that the reaction between the flux and crucible progressed when the sample was held at high temperatures for a long time and that a lower growth rate is not advantageous for (Hg,Re)1223 crystals.

Finally, we found that a final temperature of 920 °C was optimal. When the final temperature was increased to 940 °C, the crystal size decreased to less than 0.2 × 0.2 mm². However, when the temperature was reduced to 900 °C, the crystal size did not increase, suggesting that the crystal growth was terminated at 920 °C. These results suggest that the optimal temperature range for the crystal growth of (Hg,Re)1223 is approximately 920–940 °C.

As described above, the optimal conditions for each factor are within a narrow range, and large single crystals cannot be obtained if even one of these conditions is not fulfilled. Only by appropriately combining these conditions can (Hg,Re)1223 single crystals with a size of 1 × 1 mm² be obtained with good reproducibility.

## 4.   Characterization of Single Crystals

*4.1 Photographs of Hg1223 single crystals*

Fig. 4 shows a photograph of the typical as-grown (Hg,Re)1223 single crystal. The crystal size was 1.0 × 1.0 × 0.038 mm³. As is often the case with single crystals of layered materials, the crystal had flat faces, which were considered the *ab* plane. Moreover, the shape was a well-defined square, reflecting the tetragonal crystal structure of (Hg,Re)1223.

*4.2 XRD patterns and Laue photographs*

Fig. 5(a) shows the XRD pattern of a (Hg,Re)1223 crystal measured on a flat surface. Sharp peaks corresponding to the 00*l* reflections of the (Hg,Re)1223 phase are observed (the assigned 00*l* indices are shown in the figure). The *c*-axis length was estimated to be $c$ = 15.67 Å, consistent with the literature value of (Hg,Re)1223 single crystals ($c$ = 15.66 Å).[14] Peaks corresponding to other phases, such as (Hg,Re)1212 and (Hg,Re)1234, were not observed.

Fig. 5(b) shows a back-reflection Laue photograph taken on a flat surface. The Laue spots exhibited four-fold symmetry, and no splitting of spots was observed. This indicates that the sample was single-domain and confirms that the flat surface corresponds to the *ab* plane. In addition, the edges of the crystal shown in Fig. 4 correspond to the *a*(*b*)-axis directions of the crystal structure. This indicates that the in-plane crystal growth occurred along the *a*- and *b*-

axes (Cu-O bonding) directions, which is consistent with the results of a previous study.[15]

*4.3 SEM/EDX analysis*

Fig. 6 shows an SEM image of the as-grown (Hg,Re)1223 single crystal. The white spots on the crystal surface were identified as HgO, which presumably attached during the cooling process of crystal growth. These spots disappeared when the sample was annealed at 500 °C. Chemical composition analysis using EDX was performed at eight points on the sample surface, as shown in the figure. Table 3 presents the composition ratios of the constituent elements at each point and their average values when the ratio of Hg+Re was set to 1. The average composition of the sample was Hg:Re:Ba:Ca:Cu = 0.77:0.23:1.91:1.83:3.16, which is close to the target composition, i.e., 0.8:0.2:2:2:3. Moreover, the spatial variation in the composition was <2% except for Re (approximately 4%). Considering the uncertainty (error range) of the EDX measurements, these results confirm that the obtained crystals were indeed (Hg,Re)1223 and had reasonable homogeneity.

## 5. Physical Properties

*5.1 Magnetization*

Fig. 7(a) shows the temperature dependence of the magnetic susceptibility of the as-grown (Hg,Re)1223 crystal under field-cooled (FC) and zero-field-cooled (ZFC) conditions. Measurements were performed by applying a magnetic field of 3 Oe in the direction parallel to the *c*-axis. The inset shows an enlarged view around the onset of the superconducting transition. $T_c^{on}$, which is defined as the beginning of the decrease in magnetization, was 131.5 K. The superconducting transition width $\Delta T_c$, which is defined as the temperature interval from 10% to 90% of the decrease in the susceptibility, was smaller than 3 K. The volume fraction estimated from the magnetic susceptibility under the ZFC condition at 90 K was 95%.

Fig. 7(b) presents a comparison of the temperature dependence of magnetization for six crystals from the same batch. Measurements were performed under an external magnetic field of 5 Oe along the *c*-axis, and the data were normalized with respect to the value at 90 K in the ZFC condition ($M^{90\ K}$). The $T_c^{on}$ values were between 130.0 and 130.5 K for all the samples, whereas the $\Delta T_c$ values ranged from 2.5 to 6.0 K. For samples #2 and #3, a second drop of the magnetization was observed at approximately 125 K, which was attributed to the superconducting transition of the (Hg,Re)1212 or (Hg,Re)1234 phase included in the crystals. Most of the samples exhibited a sharp superconducting transition at 130 K, confirming that single-phase (Hg,Re)1223 crystals were successfully obtained.

*5.2 In-plane resistivity*

Fig. 8 shows the temperature dependence of the in-plane resistivity ($\rho_{ab}$) normalized with respect to $\rho_{ab}$ at 300 K ($\rho_{ab}^{300K}$) for the three as-grown crystals. The inset shows an enlarged view around the superconducting transition. The typical value of $\rho_{ab}^{300K}$ estimated from the three samples was 300 μΩ cm. All the samples exhibited metallic behavior; that is, the resistivity decreased as the temperature decreased. In particular, $\rho_{ab}$ exhibited a linear temperature dependence over a wide temperature range. These results indicate that the as-grown crystals in the present study were optimally doped. As shown in the inset of the figure, $\rho_{ab}$ decreased rapidly at 131, 131, and 130 K for samples #1, #2, and #3, respectively. The temperatures at which the resistivity reached zero, i.e., $T_c^{zero}$, were 128 K (#1), 126 K (#2), and 126.5 K (#3). For samples #2 and #3, the superconducting transition was broader than that of #1 and exhibited multiple steps. Because the steps occurred above the $T_c$ of Hg1212 (123K) and Hg1234 (125 K), they were probably due to the local variation in oxygen content.

The residual resistivity estimated via the linear extrapolation of $\rho_{ab}$ down to 0 K was negative for all three samples: –10 μΩ cm (#1), –60 μΩ cm (#2), and –50 μΩ cm (#3). A small residual resistivity indicates good sample quality with a low degree of impurity scattering. A negative residual resistivity was also observed for Bi2223, which contains three $CuO_2$ planes in its structural unit and has a $T_c$ higher than 100 K.[20] In contrast, for Cu oxides with one or two $CuO_2$ planes, such as La214, Y123, Bi2201, and Bi2212, the (extrapolated) residual resistivity is typically positive. The negative residual resistivity is possibly a characteristic of systems containing three or more $CuO_2$ planes and having $T_c$ values above 100 K.

*5.3 Out-of-plane resistivity*

Fig. 9 shows the temperature dependence of the out-of-plane resistivity ($\rho_c$) of the three as-grown crystals (the crystals are different from those used for the $\rho_{ab}$ measurement). The $\rho_c$ was normalized with respect to that at 300 K ($\rho_c^{300K}$), and the typical $\rho_c^{300K}$ estimated from the three samples was 130 mΩ cm. All samples exhibited metallic behavior ($d\rho_c/dT > 0$) above 230 K. Below 200 K, $\rho_c$ increased with a reduction in the temperature, and it decreased rapidly at 131 K. In the enlarged view (inset of Fig. 9), $\rho_c$ exhibits a broader superconducting transition than $\rho_{ab}$, and the $T_c^{zero}$ values are 126 K (#1), 125.5 K (#2), and 125 K (#3), which are slightly lower than those of $\rho_{ab}$. The broader transition and lower $T_c^{zero}$ may be due to the intergrowth of Hg1234 phases, as intergrowth is usually introduced along the *ab* planes and is expected to have a more significant effect on $\rho_c$ than on $\rho_{ab}$.

The resistivity anisotropy defined as $\gamma^2 = \rho_c/\rho_{ab}$ at 300 K was estimated by combining the results in Figs. 8 and 9. The obtained $\gamma^2$ was 250–650, which is consistent with a previous report ($\gamma^2 \approx 500$).[21] The uncertainty in the $\gamma^2$ value is mainly due to the sample dependence of $\rho_c^{300K}$. The stronger sample dependence of $\rho_c$ compared with $\rho_{ab}$ possibly arises from the difficulty of measurement. The $\rho_c$ measurements were conducted with current and voltage electrodes on both flat surfaces (*ab* planes) of the thin-plate samples. This method is reliable when the resistivity anisotropy is significant ($\gamma^2$ is sufficiently large), but the value of $\gamma^2 \approx 500$ indicates moderate anisotropy. This may cause differences in the current distribution depending on the positions of electrodes, possibly resulting in differences in the temperature dependence of $\rho_c$. Considering the error due to the measurement setup, we conclude that the intrinsic variation in the physical properties of the (Hg,Re)1223 crystals grown in this study was small, confirming that samples of similar quality were obtained with good reproducibility.

## 6. Summary

We established a method for growing (Hg,Re)1223 single crystals. An explosion-proof stainless-steel container controlled the high pressure of Hg vapor at high temperatures, ensuring safe crystal growth. Furthermore, good reproducibility was achieved by employing a single-step growth method using oxides as raw materials. The obtained crystals had a maximum size of $1 \times 1 \times 0.04$ mm$^3$ and exhibited a square shape reflecting the symmetry of the crystal structure. The $T_c$ of the as-grown crystals was as high as 130 K, indicating that the samples were close to the optimally doped state.

**Acknowledgment**

This work was supported by JSPS KAKENHI (JP16H03854, JP19H05823, and JP21H01377).

# Figures

## Fig. 1

(a)

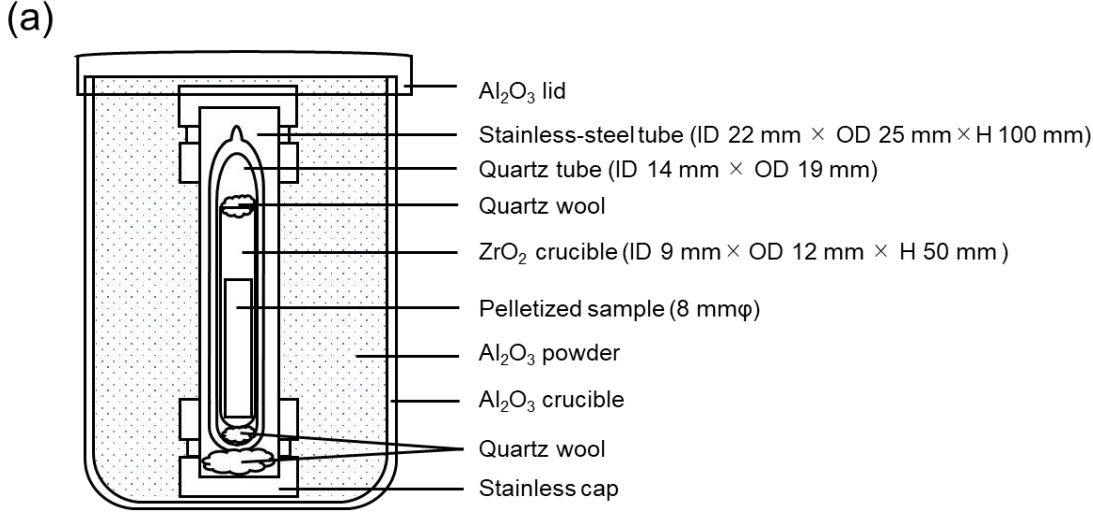

(b)

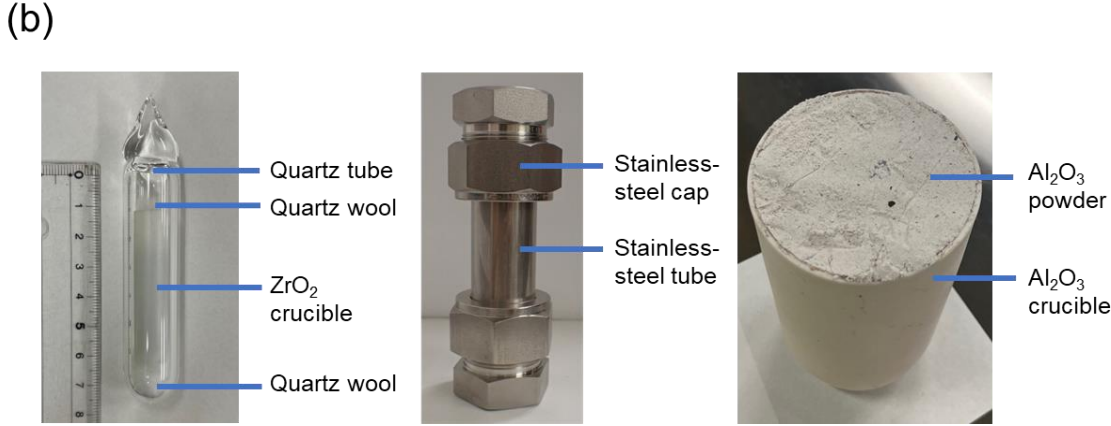

## Fig. 2

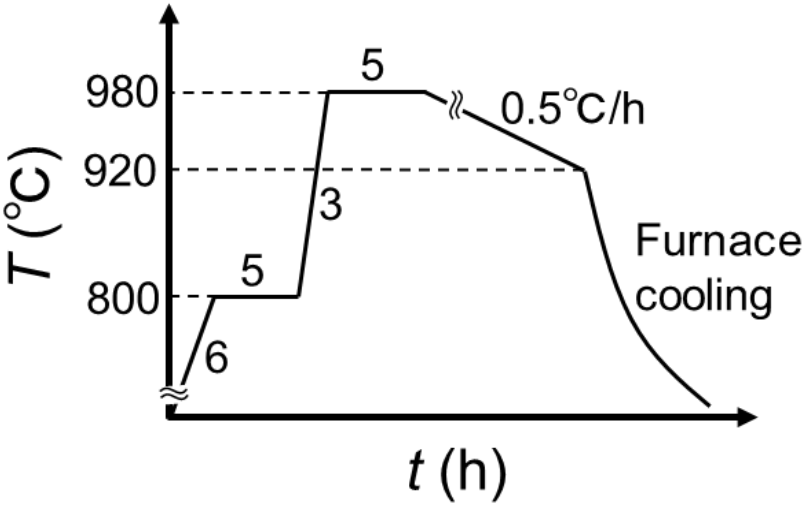

**Fig. 3**

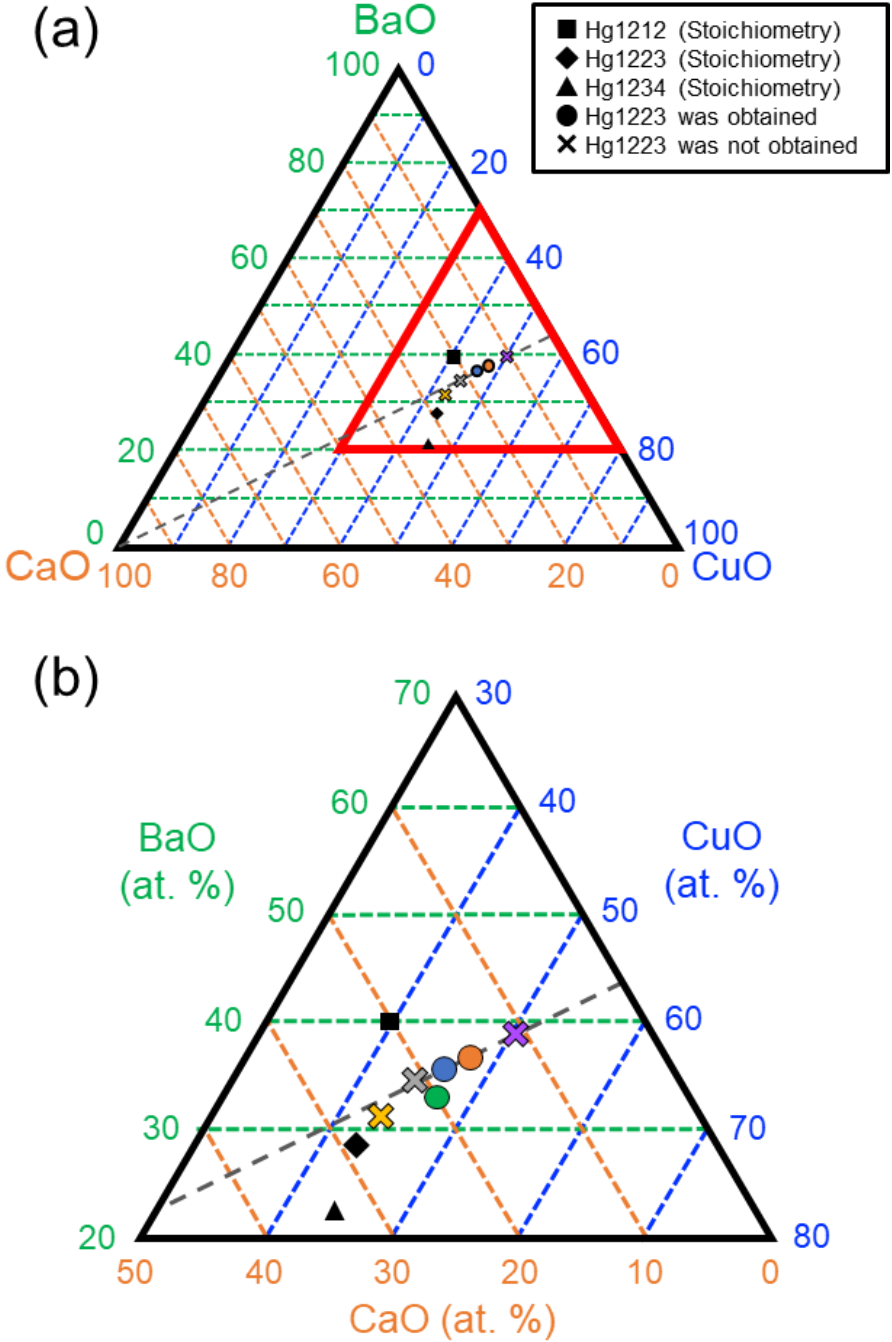

**Fig. 4**

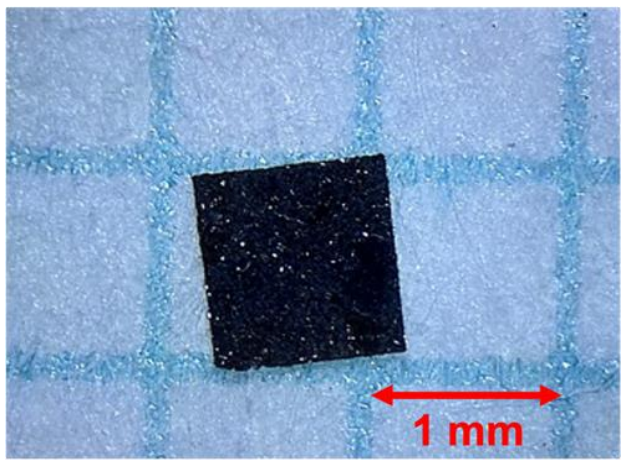

**Fig. 5**

(a)

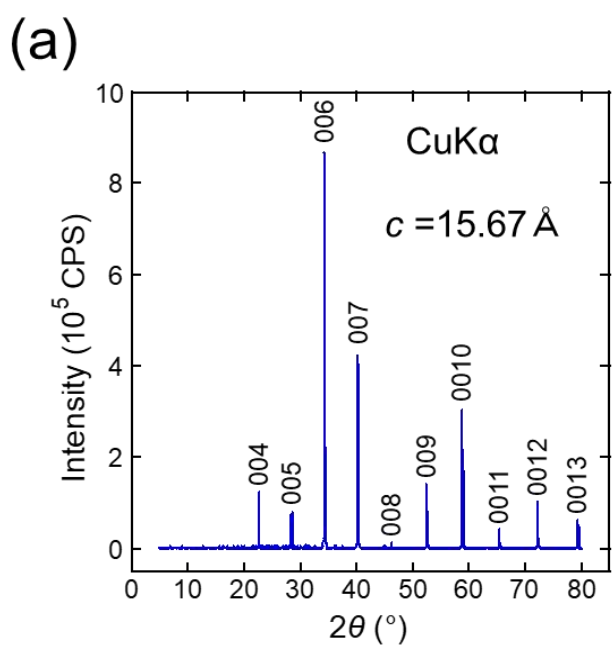

(b)

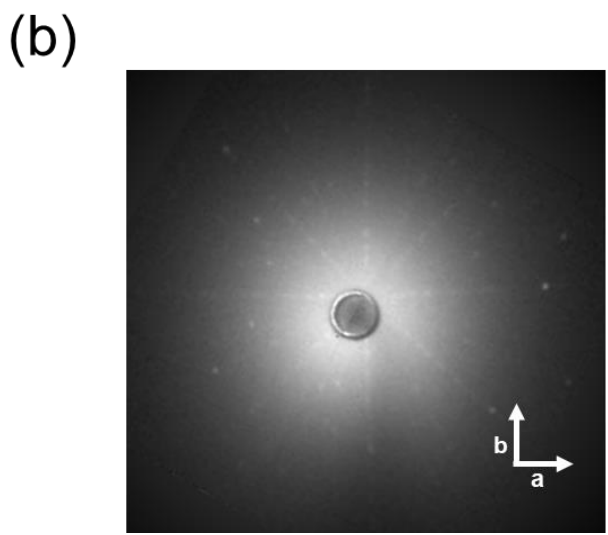

**Fig. 6**

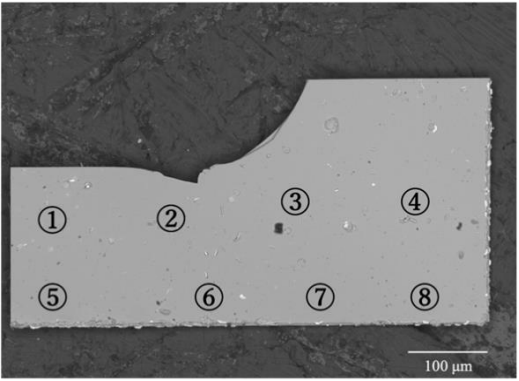

**Fig. 7**

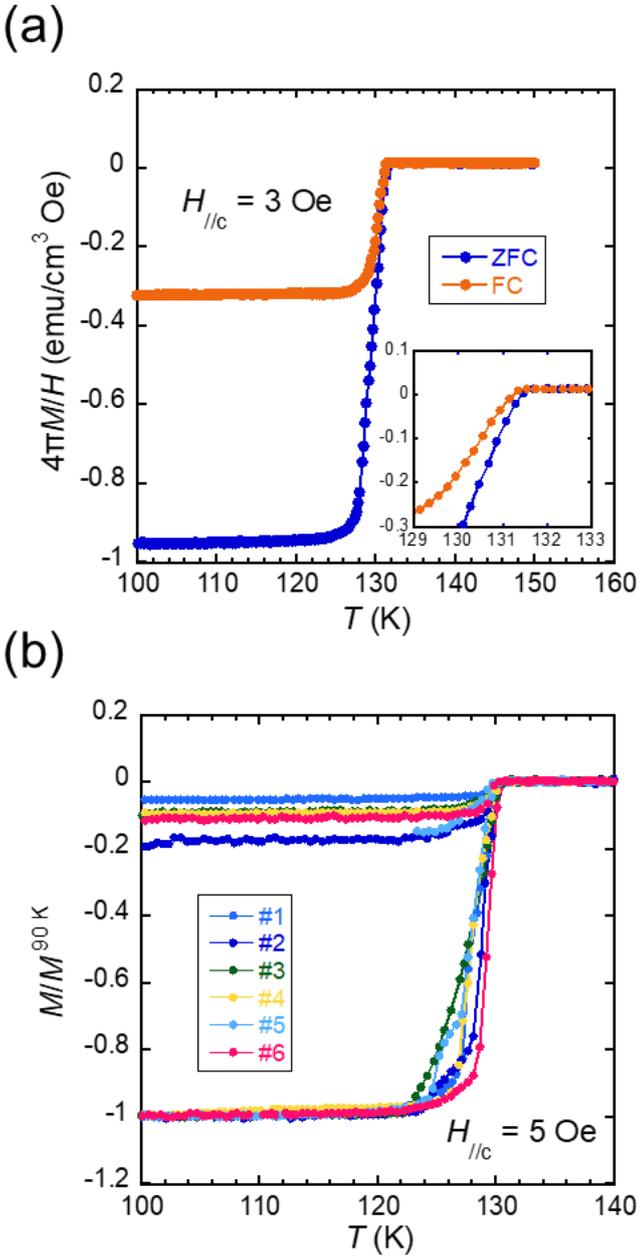

**Fig. 8**

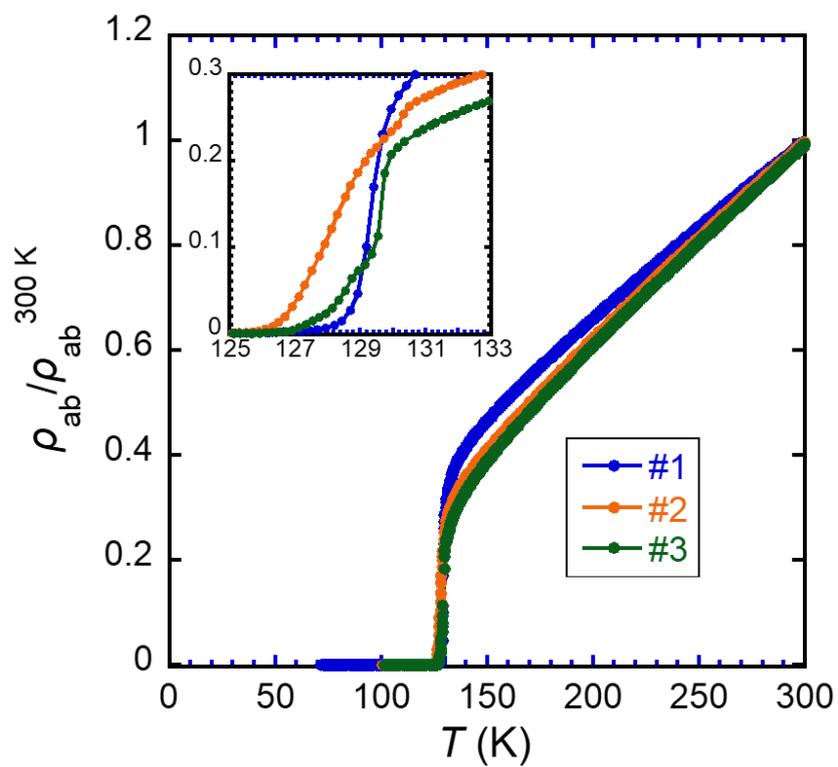

**Fig. 9**

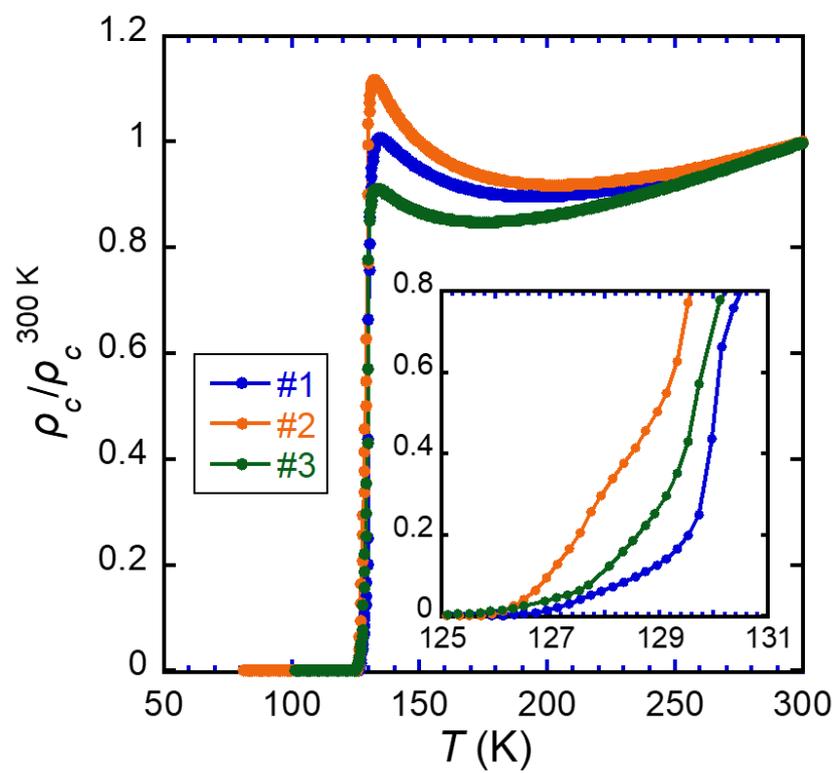

# Figure captions

**Fig. 1** (a) Schematic of the experimental setup for the growth of (Hg,Re)1223 single crystals. (b) Photographs of the experimental setup.

**Fig. 2** Optimal temperature profile for (Hg,Re)1223 crystal growth.

**Fig. 3** (a) Triangular phase diagram of BaO-CaO-CuO and (b) enlarged view of the red-framed area in (a). The square (■), diamond (◆), and triangle (▲) indicate the stoichiometric compositions for Hg1212, Hg1223, and Hg1234, respectively. Circles (●) (crosses (×)) indicate compositions for which (Hg,Re)1223 single crystals were (not) obtained. The blue circle corresponds to the optimal composition ratio in this study.

**Fig. 4** Photograph of an as-grown (Hg,Re)1223 single crystal. The blue grid indicates 1 × 1 mm$^2$.

**Fig. 5** (a) XRD pattern with the 00$l$ index and (b) back-reflection Laue photograph obtained on the flat surface of an as-grown (Hg,Re)1223 single crystal.

**Fig. 6** SEM image of the (Hg,Re)1223 single crystal. The numbers (1–8) indicate the points where the chemical compositions were analyzed using EDX.

**Fig. 7** (a) Temperature dependence of the magnetic susceptibility of an as-grown (Hg,Re)1223 single crystal under a magnetic field of 3 Oe parallel to the $c$-axis. The blue (orange) data were collected in the ZFC (FC) process. (b) Temperature dependence of the magnetization measured for six (Hg,Re)1223 single crystals under a magnetic field of 5 Oe parallel to the $c$-axis. The data were normalized with respect to the magnitude of magnetization at 90 K in ZFC process.

**Fig. 8** Temperature dependence of the in-plane resistivity ($\rho_{ab}$) for three (Hg,Re)1223 single crystals normalized with respect to $\rho_{ab}$ at 300 K ($\rho_{ab}^{300\ K}$). The inset presents an enlarged view around the superconducting transition.

**Fig. 9** Temperature dependence of the out-of-plane resistivity ($\rho_c$) for three (Hg,Re)1223 single crystals normalized with respect to $\rho_c$ at 300 K ($\rho_c^{300\ K}$). The inset presents an enlarged view

around the superconducting transition.

# Tables

**Table 1.** Single-crystal growth methods from previous studies

| Method | Raw materials ([ ] indicates a precursor) | Crucible | Encapsulation | Crystal size and $T_c$ | Heat treatment profile |
|---|---|---|---|---|---|
| Gold amalgamation[10] | HgO + BaO + CaO + CuO<br>total weight 4.06 g<br>HgO : BaO : CaO : CuO<br>1 : 2.4 : 2.4 : 3.6 | Au foil | Vacuum quartz tube<br>(ID 10 mm × OD 13 mm × H 70 mm)<br>&<br>Sealed in explosion-proof stainless-steel container | $0.34 \times 0.29 \times 0.29$ mm$^3$<br>$T_c = 120$ K (AG)<br>$T_c = 135$ K (O$_2$ 300°C 6h) | 800°C 10 h, 160°C/h |
| Flux method<br>(Bi flux)[11] | [Ba(NO$_3$)$_2$ + CaO + CuO] + HgO + Bi$_2$O$_3$<br>total weight -<br>HgO : Bi$_2$O$_3$ : Ba(NO$_3$)$_2$ : CaO : CuO<br>0.8 : 0.1 : 2 : 2 : 3 | Al$_2$O$_3$ | Vacuum quartz tube<br>(ID 9mm × H 90mm) | $300 \times 300 \times 10$ μm$^3$<br>$T_c = 130$ K (As grown) | 1020°C 10 h, 800°C 10 h, 950°C 10 min, 700°C 12 h, 5 h, 2 h |
| Flux method<br>at high pressure<br>(Pb flux)[12] | HgO + BaO + CaO + CuO<br>total weight -<br>PbO 10-20% (by weight) | Y$_2$O$_3$ | High-pressure dedicated container | $0.10 \times 0.10 \times 0.002$ mm$^3$<br>$T_c = 133$ K (As grown) | 1150°C, 1185°C, 950°C 15 min, 920°C 15 min, 45 min, 1 h, 5 h |
| Self-flux method<br>(Re dope)[13] | [BaCO$_3$ + CaCO$_3$ + CuO] + HgO + ReO$_2$<br>total weight 3 g<br>HgO : ReO$_2$ : BaCO$_3$ : CaCO$_3$ : CuO<br>0.75~0.98 : 0.25~0.02 : 2 : 1 : 2 | Al$_2$O$_3$ | Vacuum quartz tube<br>(ID 11 mm × OD 15 mm × H 90 mm) | $3 \times 2 \times 0.1$ mm$^3$<br>$T_c = 130$ K (As grown)<br>$T_c = 132$ K (O$_2$ 300°C) | 1015°C 1 h, 950°C, 1°C/h, 6 h |
| Self-flux method<br>(Re dope)[14] | [ReO$_3$ + BaCO$_3$ + CaCO$_3$ + CuO] + HgO + BaF$_2$<br>total weight 4 g<br>HgO : ReO$_3$ : BaCO$_3$ : CaCO$_3$ : CuO<br>0.75 : 0.25 : 2 : 1.5 : 2.5 | BaZrO$_3$ | Vacuum double quartz tube<br>(ID 9 mm × OD 11 mm × H 70 mm)<br>&<br>(ID 12 mm × OD 17 mm × H 110 mm) | $1 \times 1.1 \times 0.1$ mm$^3$<br>$T_c = 131$ K (As grown) | 1025°C 1 h, 975°C 0.5°C/h, 950°C, 920°C, 3 h, 0.5 h, 1.5 h |
| Self-flux method[15] | HgO + BaO + CaO + CuO<br>total weight 4.1 g<br>HgO : BaO : CaO : CuO<br>1 : 2.2 : 2.2 : 3.6 | Al$_2$O$_3$ | Vacuum quartz tube<br>(ID 13 mm × H 70 mm) | $0.4 \times 0.5 \times 0.3$ mm$^3$<br>$T_c = 105$ K (As grown)<br>$T_c = 133$ K (O$_2$ 325°C) | 860°C 3 h, 725°C 5°C/h, 600°C, 5-24 h, 300°C/h |
| Self-flux method at<br>high pressure[16] | [Ba(NO$_3$)$_2$ + CuO] + CaO + HgO<br>total weight –<br>HgO : Ba(NO$_3$)$_2$ : CaO : CuO<br>1+δ : 2 : 2 : 3 | ZrO$_2$ | Vacuum quartz tube<br>(3 mm thickness) &<br>Applied N$_2$ counter pressure<br>(Max 500 bar) | $2 \times 1 \times 0.3$ mm$^3$<br>$T_c = 127$ K (As grown) | 1040°C 10 h, 2°C/h, 990°C, 830°C 20 h, 10 h |

**Table 2.** Results of compositional analysis of (Hg,Re)1223 single crystal by EDX

| Element | ① | ② | ③ | ④ | ⑤ | ⑥ | ⑦ | ⑧ | Average |
|---|---|---|---|---|---|---|---|---|---|
| Hg | 0.766 | 0.772 | 0.758 | 0.771 | 0.774 | 0.773 | 0.764 | 0.767 | 0.768 |
| Re | 0.234 | 0.228 | 0.242 | 0.229 | 0.226 | 0.227 | 0.236 | 0.233 | 0.232 |
| Ba | 1.91 | 1.92 | 1.90 | 1.91 | 1.91 | 1.90 | 1.91 | 1.92 | 1.91 |
| Ca | 1.80 | 1.86 | 1.83 | 1.85 | 1.84 | 1.86 | 1.82 | 1.81 | 1.83 |
| Cu | 3.15 | 3.18 | 3.15 | 3.24 | 3.17 | 3.13 | 3.12 | 3.12 | 3.16 |